\documentclass[reprint,
superscriptaddress,
longbibliography,
amsmath,amssymb,
aps,
prl,
]{revtex4-1}

\usepackage{graphicx}
\usepackage{dcolumn}
\usepackage{bm}

\usepackage[colorlinks,linkcolor=blue,citecolor=blue,urlcolor=blue]{hyperref}

\usepackage{textcomp} 

\usepackage{gensymb}

\usepackage{color}

\begin{document}

\title{Strong Reverse Saturation and Fast-Light in Ruby}

\author{Akbar Safari}
\email[]{Akbar.Safari@gmail.com}
\affiliation{Department of Physics, University of Ottawa, Ottawa, ON, K1N 6N5, Canada.}
\affiliation{Department of Physics, University of Wisconsin-Madison, Madison, WI 53706, USA.}

\author{Cara Selvarajah}
\affiliation{Department of Physics, University of Ottawa, Ottawa, ON, K1N 6N5, Canada.}

\author{Jenine Evans}
\affiliation{Department of Physics, University of Ottawa, Ottawa, ON, K1N 6N5, Canada.}

\author{Jeremy Upham}
\affiliation{Department of Physics, University of Ottawa, Ottawa, ON, K1N 6N5, Canada.}

\author{Robert W. Boyd}
\affiliation{Department of Physics, University of Ottawa, Ottawa, ON, K1N 6N5, Canada.}
\affiliation{Institute of Optics, University of Rochester, Rochester, New York, 14627, USA.}

\vspace{3mm}

\date{\today}

\begin{abstract}
We observe a strong reverse saturation of absorption in ruby at a wavelength of 473\,nm. With an intensity-modulated laser, we observe that the peaks of the pulses appear more than a hundred microseconds earlier than the reference signal. A theoretical model based on coherent population oscillation would suggest a fast-light effect with an extremely large and negative group index of $-(1.7\pm0.1)\times 10^6$. We propose that this pulse advancement can also be described by time-dependent absorption of ruby. Our study helps to understand the nature of the fast- and slow-light effects in transition-metal-doped crystals such as ruby and alexandrite.
\end{abstract}

\maketitle
Saturation of absorption is a well-understood nonlinear optical process in which an optical driving field passing through an absorptive material experiences a decrease in absorptivity as the intensity of the field increases\,\cite{BoydNL}. This process is typical of most materials and is frequently employed for passively mode-locking or Q-switching lasers\,\cite{Siegman}.

However, there are conditions under which materials exhibit an increase in absorption with higher intensity at specific wavelengths. This response, called reverse saturation of absorption (RSA), requires particular conditions, including a more than two-level system and for an excited state to have a larger absorption cross-section than the ground state\,\cite{Prior}. In addition, neither the first nor the second excited states should decay to other levels thereby trapping the population. Moreover, the incident light should saturate, or partially saturate, the first transition only. This condition can be achieved easily when the lifetime of the first excited state is much longer than that of the second excited state. Experimental investigation of RSA, particularly their temporal dynamics, will provide insight into their feasibility for power limiting\,\cite{TUTT1993} and fast-light applications in gravitational wave detection\,\cite{Shahriar2008}, optical gyroscopes\,\cite{Shahriar2007,Salit2007}, and more\,\cite{Boyd2009,MilonniFastLight,Akulshin}.

Here, we show that ruby at room temperature exhibits a strong reverse saturation of absorption at the wavelength of 473\,nm. Consequently, we observe what could appear to be a fast-light effect, where the peak of an intensity-modulated signal passing through the crystal reaches the detector earlier than that without the ruby crystal. This phenomenon had been explained earlier based on coherent population oscillation and hole burning \cite{BigelowPRL, BigelowSci,Malcuit84}. While this model continues to be debated\,\cite{BigelowPRL,BigelowSci,Piredda2007,Barker2014,Kozlov2014,Kozlov2006,Segard2008,Segard2010,SegardPRA}, we consider a simple model based on rate-equations that explains the seemingly fast-light effect without the need of hole or anti-hole burning. In the following, first, we confirm that ruby exhibits RSA. Then, we use rate equations to find the time-dependent population of the ground and excited states, and consequently, the time-dependent absorption of ruby. Finally, we show that the advancement of the peak intensities can be explained by time-dependent absorption. 

\begin{figure}[t]                    
\begin{center}
\includegraphics[width=8.6cm]{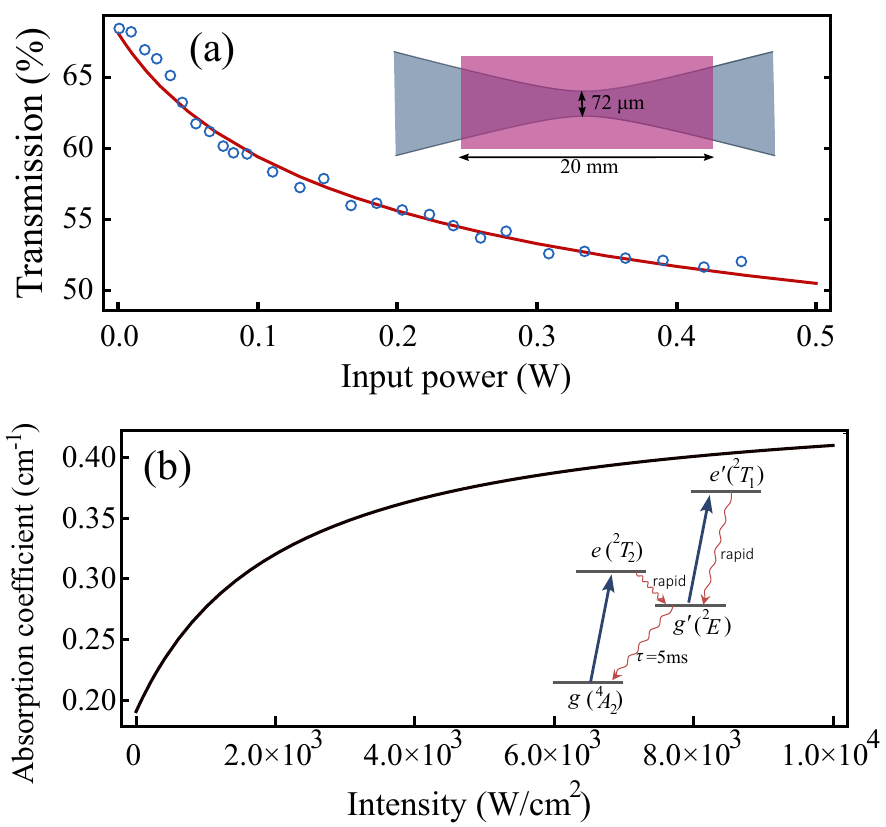}
\end{center}
\caption{\textbf{Reverse saturation of absorption.} (a) Experimental (open circles) and theoretical (solid line) transmission as a function of the input laser power for a CW (unmodulated) beam. The inset shows the beam profile and the position of ruby. (b) Theoretical absorption coefficient from Eq.~\eqref{Alpha}. Inset: The relevant energy levels of ruby. The absorption cross-section of the second transition, $g' \rightarrow e'$, is larger than that of the first transition, $g \rightarrow e$, which leads to reverse saturation of absorption.}
\label{RSA} 
\end{figure}

The reverse saturation of absorption is observed by measuring the transmission of a continuous-wave (CW) laser at wavelength 473\,nm with a maximum power of 500\,mW. To achieve the desired intensities, the laser is focused to a beam waist of 36\,$\mu$m at the center of the ruby crystal of length 20\,mm. The results, shown in Fig.~\ref{RSA}(a), show a clear reduction of the transmission as the laser power increases. The crystal is placed at a small angle to avoid any issues with the back reflection of the laser from the crystal faces. The Fresnel reflections are considered in the calculation of absorption. We observe that the transmission through the ruby crystal decreases from 68\% in the linear regime, to 52\% in the nonlinear (high power) regime.

The relevant energy levels of ruby are drawn in the inset of Fig.~\ref{RSA}(b). The laser excites the electrons from the ground state $g$ to the excited state $e$. The excited electrons relax very rapidly to the meta-stable state $g'$ by emitting phonons. These three levels are typically enough to describe the interaction of the Cr$^{3+}$ ions with a green laser, for example. However, a blue laser at 473\,nm, can excite the electrons from the meta-stable state to the second excited state, $e'$. Because the absorption cross-section of this transition, $\sigma_2$, is larger than that of the first transition, $\sigma_1$, the overall absorption rate increases with the intensity of the laser (as shown in Fig.~\ref{RSA}(b)).  

Since states $e$ and $e'$ have extremely short lifetimes, their populations are negligible and can be omitted in the rate equation. Therefore, the population density of the ground state $N_g$ follows
\begin{eqnarray}                     
\frac{dN_g}{dt} = -\frac{I}{\hbar \omega} \sigma_1 N_g + \frac{N-N_g}{\tau}
\label{Ng} 
\end{eqnarray}
where, $I$ is the intensity of the laser with photon energy $\hbar \omega$, and $N-N_g$ is the population density of the meta-stable state $g'$, with lifetime $\tau=$5\,ms \cite{Siegman}. $\sigma_1$ is the absorption cross-section of the first transition from $g$ to $e$. The density of the Cr$^{3+}$ ions in the ruby crystal is $N\approx4.75\times 10^{18}$ cm$^{-3}$. For a CW excitation, Eq.~\eqref{Ng} is solved in steady-state condition $(dN_g/dt=0)$. Then, the absorption coefficient is found from
\begin{eqnarray}                     
\alpha = & \sigma_1 N_g + \sigma_3 (N-N_g) \\
=& \frac{N}{1+I/I_s} (\sigma_1 - \sigma_3) + \sigma_3 N,
\label{Alpha} 
\end{eqnarray}
where $\sigma_3$ is the absorption cross-section of the second transition, and $I_s=\hbar \omega /\sigma_1 \tau$ is the saturation intensity. We note that since the populations of levels $e$ and $e'$ are negligible, implementing the degeneracy factors of the states will not affect Eq.~\eqref{Alpha}. Figure~\ref{RSA}(b) plots the absorption coefficient from Eq.~\eqref{Alpha} as a function of intensity. 

In order to test the validity of Eq.~\eqref{Alpha} with experimental data, we simulate the propagation of the laser through the ruby crystal and find the theoretical transmission as a function of the input power. We use the same beam profile as measured in the experiment (inset of Fig.~\ref{RSA}(a)). Since the local intensity and the absorption coefficient are interdependent, we adopt an iterative approach to calculate the intensity of the laser as it propagates through the crystal. The results, shown as a solid, red line in Fig.~\ref{RSA}(a), exhibit an excellent agreement with the experimental data. Therefore, we extract the values of the absorption cross sections to be $\sigma_1=4.0\times10^{-20}$\,cm$^2$ and $\sigma_3=9.6\times10^{-20}$\,cm$^2$. 
It is because the absorption cross-section of the second transition is nearly 2.5 times larger than that of the ground state at this wavelength that enables the clear observation of reverse saturation of absorption in this experiment.

\begin{figure}[t!]                    
\begin{center}
\includegraphics[width=8.6cm]{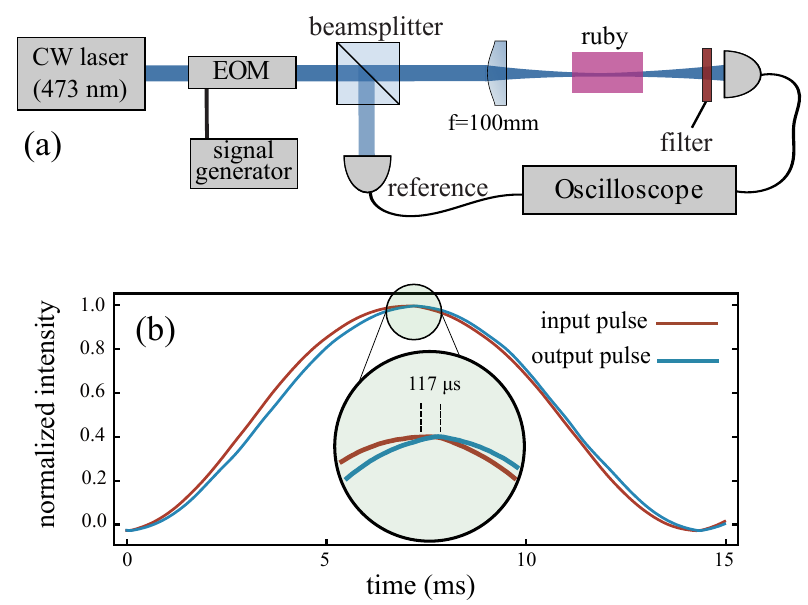} 
\end{center}
\caption{\textbf{Fast-light experiment}. (a) The experimental setup. An electro-optic modulator (EOM) is used to imprint a weak sinusoidal intensity modulation on the laser beam. The laser is focused in the ruby crystal. A spectral filter is used to block the fluorescence at 694\,nm. (b) Upon propagation in ruby, the peak of the weakly modulated signal advances in time for approximately 117\,$\mu$s compared to the reference signal.}
\label{Setup} 
\end{figure}

Next, we investigate the fast-light effect and examine how the temporal variations of a modulated beam are altered upon traveling through the reverse saturated medium. Following refs.\,\cite{BigelowPRL,BigelowSci}, we send the laser beam through an electro-optic device which imposes a 10\% intensity modulation to the laser beam, Fig.~\ref{Setup}(a). The electro-optic modulator is fed by a sinusoidal signal at frequency $\Omega/2\pi=70$\,Hz. The beam is focused to a waist of 36\,$\mu$m at the center of the ruby crystal. A spectral filter is used to filter out the fluorescence at 694\,nm. The transmitted and reference signals are detected and compared on an oscilloscope. When operating at intensities where RSA is clearly visible, the peaks of the modulated signal appear to advance relative to the reference. Figure~\ref{Setup}(b) shows $117\pm 6\,\mu$s pulse advancement for an average input power of 450\,mW.

These results appear to be consistent with similar experiments in ruby and alexandrite\,\cite{BigelowPRL,BigelowSci}, where slow- and fast-light effects were reported and attributed to \textit{coherent population oscillation}. Were the same reasoning used here, the RSA would lead to a spectral \textit{hill} or anti-hole and the corresponding advancement of the pulse peak would indicate fast-light with a group index of $-(1.7\pm0.1)\times 10^6$. Here we examine another hypothesis to describe these results: a time-dependence to ruby's response to the modulated signal.

\begin{figure}[t!]                    
\begin{center}
\includegraphics[width=8.6cm]{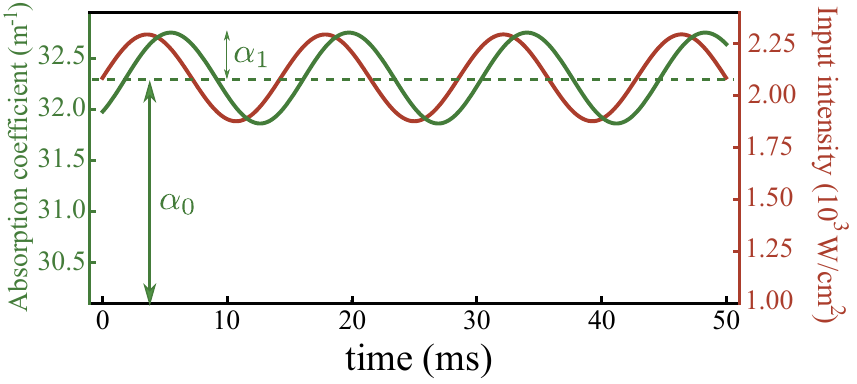}
\end{center}
\caption{\textbf{Ruby time-dependent response}. Modulated input intensity (right vertical axis) and the corresponding time-dependent absorption coefficient (left vertical axis). When the frequency of modulation is small compared to $1/\tau$, the absorption oscillates with the driving intensity, however, with a time difference due to the finite lifetime of the meta-stable state. As the frequency of modulation increases, the oscillation amplitude of the absorption coefficient, $\alpha_1$, decreases.}
\label{Fig3} 
\end{figure} 

For an intensity-modulated laser beam, the input intensity to the ruby crystal can be written as
\begin{eqnarray}                     
I(t) = I_0 + I_1 e^{-i\Omega t} + \rm{c.c.},
\label{Int} 
\end{eqnarray}
where, $\Omega$ is the angular frequency of the modulation, $I_0$ is the average intensity, and $I_1$ is the amplitude of the modulation. In Refs.~\cite{BigelowPRL, Piredda2007, Malcuit84}, in order to explain an anti-hole in the spectrum, the population of the ground state and the absorption were written as $N(t) = N_0 + N_1 e^{-i\Omega t} +$c.c., and $\alpha(t) = \alpha_0 + \alpha_1 e^{-i\Omega t} + $c.c., respectively. In other words, it was assumed that the population and the absorption follow the same time dependence as the driving intensity. Therefore, the plot of $\alpha_0+\alpha_1$ as a function of the modulation frequency exhibited a narrow peak, which was interpreted as an anti-hole. 

However, the total absorption of ruby is time-dependent, with $\alpha_0+\alpha_1$ merely showing the maximum absorption (see Fig.~\ref{Fig3}). When the period of the modulation is long enough compared to the lifetime of the meta-stable state $\tau=5$\,ms, the population can vary, following the driving intensity. Because the absorption cross-section of the meta-stable state is larger than that of the ground state, as the population of the meta-stable state increases, the total absorption, $\alpha(t)$, increases as well. Therefore, $\alpha_1$ is maximum at slow modulations. As the frequency of the modulation increases, the amplitude of $\alpha_1$ decreases because the population cannot keep up with the rapidly changing intensity. Therefore, $\alpha_1$ decreases at fast modulations.

\begin{figure}[t]                    
\begin{center}
\includegraphics[width=8.6cm]{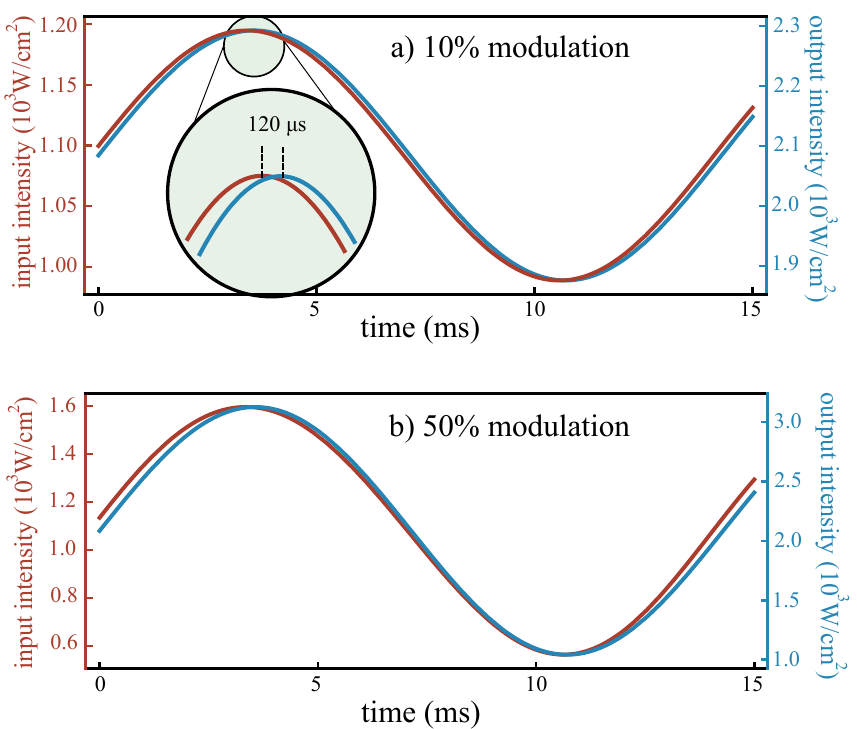}
\end{center}
\caption{\textbf{Theoretical pulse advancement.} The input and output (after propagation through ruby) intensity as a function of time for (a) weak modulation and (b) strong modulation calculated from time-dependent absorption of ruby. When the modulation is weak, the signal is advanced uniformly. With strong modulations, the output signal is deformed notably. }
\label{Fig4} 
\end{figure} 

Given that the driving intensity has the time-dependent form given by Eq.~\eqref{Int}, the population and absorption may lag behind the driving intensity as the lifetime of the meta-stable state is relatively long. Therefore, we numerically solve the rate equation in Eq.~\eqref{Ng} with the modulated intensity to find the population as a function of time. The absorption is calculated by $\alpha(t)=\sigma_1 N_g(t) + \sigma_3 (N-N_g(t))$. Figure~\ref{Fig3} plots the driving intensity and the calculated absorption as a function of time. There is clearly a time difference between the maximum of absorption and the maximum of the input intensity due to the finite lifetime of the meta-stable state. The time-dependent absorption lagging behind the modulated intensity reshapes the temporal form of the intensity as measured at the output of the crystal.

Accounting for this time-dependent intensity and absorption, we simulate the form of the output intensity as a function of time (Fig.~\ref{Fig4}). We observe that the peak of the output intensity appears earlier than the peak of the input intensity, resembling the experimental results of Fig.~\ref{Setup}(b). We calculate a time-advancement of $120\,\mu$s, equivalent to a group index of $-1.8 \times 10^6$, which are in excellent agreement with the experimental observations. For simplicity, we have ignored any change of absorption along the ruby crystal due to the change of intensity through either diffraction or absorption. A more accurate quantitative result can be obtained by employing an iterative approach. 

\begin{figure}[t]                    
\begin{center}
\includegraphics[width=8.6cm]{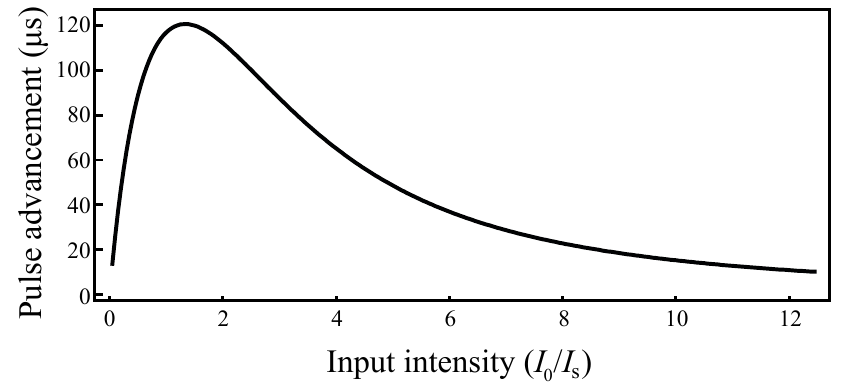}
\end{center}
\caption{\textbf{Theoretical pulse advancement as a function of intensity}. The pulse advancement reaches the maximum at intensities about the saturation intensity. As the laser intensity increases further, the effect decreases gradually.}
\label{Fig5} 
\end{figure}

Having such a theoretical model at hand, one can easily calculate other responses of interest. For example, Fig.~\ref{Fig4}(b) shows that the output intensity is deformed when the amplitude of the modulation is large. Therefore, the pulse advancement will not be uniform across the entire period. This deformation can also be seen in our experimental results in Fig.~\ref{RSA}(b). We also calculate the pulse advancement as a function of the input intensity, $I_0$ (Fig.~\ref{Fig5}). Interestingly, the time-advancement reaches a maximum for intensities around the saturation intensity $I_s=\hbar \omega /\sigma_1 \tau$, and decreases gradually for higher intensities. 

In summary, we observed a strong reverse saturation of absorption in ruby at a wavelength of $473$\,nm, which is a consequence of the excited state having an absorption cross-section larger than the ground state. Although the requirements for RSA are stringent, we showed that ruby can exhibit a strong RSA on the blue side of the visible spectrum. This effect has been used to demonstrate a pulse advancement which could indicate a fast-light effect with a very large and negative group index. We showed that this observation can also be explained very well based on the sluggish response of ruby, consistent with the previous theories \cite{Kozlov2006, Segard2008, Segard2010, SegardPRA}. Investigating the relative validity of the sluggish time-dependent absorption theory versus the coherent population oscillation theory will require further experimental investigations. However, this experimental investigation does serve to better understand the nature of the pulse-delay and pulse-advancement in transition-metal-doped crystals, which is crucial for applications in optical delay lines and optical memories\,\cite{Boyd2009, MilonniFastLight, NeveuPRL2017}, optical gyroscopes\,\cite{Shahriar2007, Salit2007}, and photon-drag\,\cite{SonjaSci2011, SafariPRL2016}. 

\subsection{Acknowledgment}
This work was supported by the Canada Excellence Research Chairs program and the National Science and Engineering Research Council of Canada (NSERC). 

\bibliography{Ref}

\end{document}